\documentclass{WileyMSP-template}

\usepackage{hyperref}
\usepackage{pdfpages}

\hypersetup{colorlinks=true,
    linkcolor=blue,
    citecolor=blue, urlcolor=blue}

\begin{document}

\pagestyle{fancy}
\rhead{}
\title{Unraveling p-type and n-type interfaces in Superconducting\\ Infinite-Layer Nickelate thin films}

\maketitle

% Author: Please give full first and last names for authors and include * after the name of all corresponding authors

\author{Aravind Raji*,}
\author{Araceli Guti{\'e}rrez-Llorente,}
\author{Dongxin Zhang,}
\author{Xiaoyan Li,}
\author{Manuel Bibes,}
\author{Lucia Iglesias,}
\author{Jean-Pascal Rueff*,}
\author{Alexandre Gloter*}

% Affiliations: Please provide adacemic titles (Prof. or Dr.) for all authors where applicable, and include an institutional email address for all corresponding authors

\begin{affiliations}
Aravind Raji\\
Laboratoire de Physique des Solides, CNRS, Universit\'{e} Paris-Saclay, 91405 Orsay, France\\
Synchrotron SOLEIL, L’Orme des Merisiers, BP 48 St Aubin, 91192 Gif sur Yvette, France\\
Email Address: aravind.raji@universite-paris-saclay.fr

\hfill \break
Araceli Guti{\'e}rrez-Llorente\\
Universidad Rey Juan Carlos, Escuela Superior de Ciencias Experimentales y Tecnolog\'{i}a, Madrid 28933, Spain\\
\hfill \break
Dongxin Zhang, Manuel Bibes, Lucia Iglesias\\
Laboratoire Albert Fert, CNRS, Thales, Universit\'{e} Paris-Saclay, 91767 Palaiseau, France\\

\hfill \break
Xiaoyan Li\\
Laboratoire de Physique des Solides, CNRS, Universit\'{e} Paris-Saclay, 91405 Orsay, France\\
\hfill \break
Jean-Pascal Rueff\\
Synchrotron SOLEIL, L’Orme des Merisiers, BP 48 St Aubin, 91192 Gif sur Yvette, France\\
LCPMR, Sorbonne Universit\'{e}, CNRS, 75005 Paris, France\\
Email Address: jean-pascal.rueff@synchrotron-soleil.fr

\hfill \break
Alexandre Gloter\\
Laboratoire de Physique des Solides, CNRS, Universit\'{e} Paris-Saclay, 91405 Orsay, France\\
Email Address: alexandre.gloter@universite-paris-saclay.fr
\end{affiliations}

% Keywords: Please provide a minimum of three and a maximum of seven keywords, separated by commas


\justifying
% Abstract should be written in the present tense and impersonal style (i.e., avoid we), and be at most 200 words long
\begin{abstract}
After decades of research, superconductivity was finally found in nickel-based analogs of superconducting cuprates, with infinite-layer (IL) structure. These results are so far restricted to thin films in the case of IL-nickelates. Therefore, the nature of the interface with the substrate, and how it couples with the thin film properties is still an open question. Here, using scanning transmission electron microscopy (STEM)- electron energy loss spectroscopy (EELS) and four-dimensional (4D)-STEM, a novel chemically sharp p-type interface is observed in a series of superconducting IL-praseodymium nickelate samples, and a comparative study is carried out with the previously reported n-type interface obtained in other samples. Both interfaces have strong differences, with the p-type interface being highly polar. In combination with ab-initio calculations, we find that the influence of the interface on the electronic structure is local, and does not extend beyond 2-3 unit cells into the thin film. This decouples the direct influence of the interface in driving the superconductivity, and indicates that the IL-nickelate thin films do not have a universal interface model. Insights into the spatial hole-distribution in SC samples, provided by monochromated EELS and total reflection-hard x-ray photoemission spectroscopy, suggest that this particular distribution might be directly influencing superconductivity.  
\end{abstract}

% Text: Please use section headings and subheadings as specified below. For communications, all section headings apart from Experimental Section should be removed
% Please make the first reference to a display item bold: \textbf{Figure 1}
% Do not abbreviate Figure, Equation, etc.; display items are always singular, i.e., Figure 1 and 2.
% Equations are always singular, i.e., Equation 1 and 2, and should be inserted using the {equation} environment, not as graphics
% Please do not use footnotes in the text, additional information can be added to the Reference list.

\section{Introduction}
The synthesis of superconducting (SC) nickelate thin films \cite{li2019superconductivity} with infinite-layer (IL) ABO$_2$ structure is one of the most significant milestones in recent years in the field of unconventional superconductivity. A puzzling fact however is the absence of superconductivity in bulk IL-nickelate compounds with the same nominal composition \cite{li2020absence,wang:20,Puphal:21}, perhaps due to the presence of extended defects. Yet, the properties of oxide thin films are known to strongly depend on the interface of the film with the substrate, so one may wonder whether there is not something specific to thin films in realizing the superconducting state. Indeed, in oxide thin film heterostructures the interface often determines the physics of the samples \cite{hwang2012emergent,mannhart2010oxide,zubko:11} including superconductivity.\cite{gariglio:11,gariglio:15}. Although, there are recent reports of superconductivity on IL-nickelate films grown on (LaAlO$_{3}$)$_{0.3}$(Sr$_{2}$TaAlO$_{6}$)$_{0.7}$ (LSAT) substrates \cite{lee2023linear,ren2023possible}, most results are on films grown on SrTiO$_3$ (STO) substrates. Intrhinguinly, superconducting IL-nickelate films grown on STO present only one particular interface termination type to date \cite{wang2023experimental}. This raises questions on structural differences between bulk and films, and on the role of the interface with the substrate in achieving the zero-resistance state. 

Regarding the interface aspect, previous scanning transmission electron microscopy (STEM) experiments reported one interface model in doped and un-doped thin films \cite{goodge2023resolving, krieger2022synthesis,osada2020superconducting} and in superlattices \cite{yang2023thickness}. This was represented by a B-site intermixed perovskite layer, considered as a buffer layer that directly quenched the two-dimensional electron gas (2DEG) expected to form at the otherwise highly polar NdNiO$_{2}$/STO interface \cite{goodge2023resolving}. This B-site intermixed interface, has a reduced polarity mismatch coming from NdO(+1 charge)-(Ni/Ti)O$_{2}$(0 charge) layer sequence, typical of a n-type interface \cite{geisler2020fundamental}. There, it was concluded from several SC and non-SC samples that this interface model is somewhat universal, and it decoupled the possible interfacial 2DEG from superconductivity in IL nickelates. Such a concept of the universal B-site intermixed interface in IL nickelates was challenged by the recent reports on synthesizing SC IL nickelate thin films with a chemically sharp interface by molecular beam epitaxy that avoided such a Ni/Ti intermixing \cite{li2023synthesis,yan2024superconductivity}. In such chemically sharp interfaces the polarity mismatch could be stronger, and how it influences the crystal and electronic structure, and thereby the superconductivity is still an open question. A comparative study on these different interfaces by advanced characterization techniques possessing high spatial and energy resolution that can isolate the interface structural and electronic effects is so far lacking in the domain of nickelate superconductivity. Notably, ab-initio calculations on IL nickelates with different substrates previously discussed the more probable formation of interfaces with n-type band alignment on STO \cite{geisler2020fundamental}, and with p-type band alignment on a NdGaO$_{3}$ substrate, here driven by +3/-1 interface polarity \cite{geisler2021correlated}. As discussed before, the n-type interface was previously reported, but so far there are no experimental reports of a p-type interface in IL nickelates. In this direction, we investigate different praseodymium IL nickelate samples, SC and non-SC, all being 20$\%$ Sr doped. By topotactic CaH$_{2}$ reduction of the parent perovskite, we obtain the SC Pr$_{0.8}$Sr$_{0.2}$NiO$_{2}$(PSNO)/STO system, that structurally involves a direct transition from the perovskite substrate to the IL without displaying a nickelate perovskite unit cell (u.c.) at the interface \cite{gutierrez2024towards}. Through a spectro-microscopic analysis with STEM, 4D-STEM, and monochromated electron energy loss spectroscopy (EELS) in combination with ab-initio calculations, we identify a novel chemically sharp p-type interface, in contrast to the B-site intermixed (n-type) interface that we obtained in other samples reduced in-situ by aluminium.

Apart from the interface aspect, to date, there are no studies on the spatial hole distribution in SC IL nickelates, notwithstanding the fact that superconductivity arises in nickelates upon hole-doping. Competing orders with superconductivity, such as the 3$a_{o}$ charge order \cite{krieger2022charge,tam2022charge,rossi2022broken}, was found to be emerging from a peculiar oxygen vacancy ordering in IL nickelates \cite{raji2023charge,raji2024valence,parzyck2024absence}. On similar grounds, there are theoretical studies on how doping suppresses this electronic instability, and paves the way for phonon-mediated superconductivity from bond disproportionation in nickelates \cite{alvarez2022charge}. To address this, we also carried out a spatial mapping of the hole-doping in the SC PSNO thin-film using depth resolved total reflection hard x-ray photoemission spectroscopy (TR-HAXPES), that also sheds some light on the IL nickelate superconductivity.               

\section{Results}
\subsection{The Interface dichotomy}
\begin{figure*}[!ht]
  \centering
  %[width=7in]
  \includegraphics[width=\textwidth]{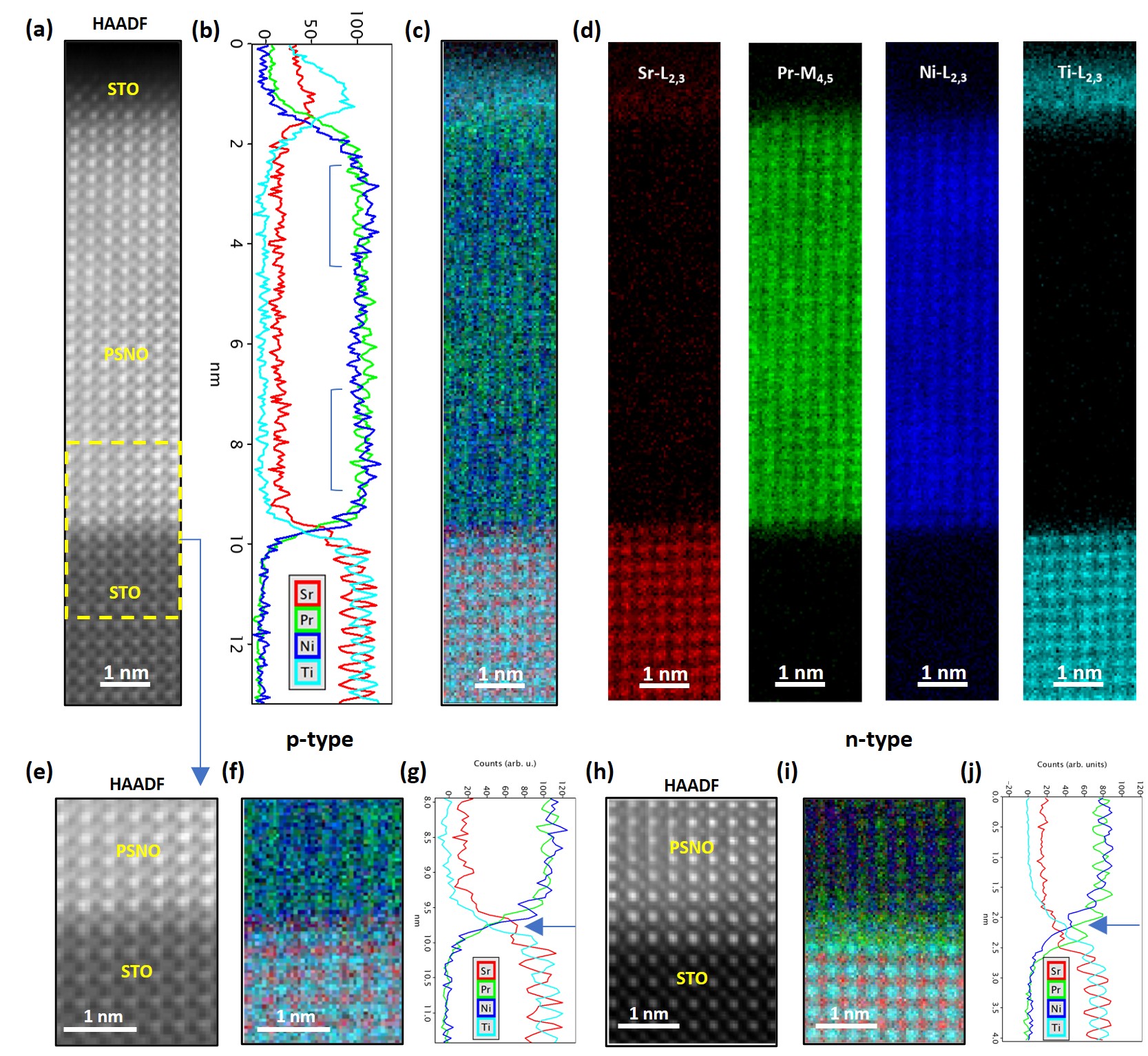}
  \caption{ STEM-EELS interface elemental analysis of SC STO-PSNO-STO and Al-PSNO-STO samples. (a) HAADF image of STO-PSNO-STO thin film. (b) The EELS element profile of maps is shown in (c,d). A slight Ni/Pr off-ratio is seen in the top and bottom of the thin-film, as highlighted. (c) Combined element map of all four Sr, Pr, Ni and Ti elements. (d) Separated element maps from the Sr-L$_{2,3}$, Pr-M$_{4,5}$, Ni-L$_{2,3}$, and Ti-L$_{2,3}$ edges. (e-g) The novel p-type SrO terminated interface in SC STO-PSNO-STO sample; (e) The magnified HAADF image at the bottom interface, (f) The combined element map of this region, (g) The element map intensity profile showing Sr-termination. (h-j) The n-type interface in Al-PSNO-STO sample;  (h) HAADF image of Al-PSNO-STO thin film's substrate interface,  (i) EELS combined element map in this region, (j) The element map intensity profile showing the Ni/Ti mixed interface model suggested in \cite{goodge2023resolving}, which we identify as n-type. 
}
  %\label{fig:boat1}
  \label{fig_STEM_EELS}
\end{figure*}

Here, we report the new p-type interface model in the SC PSNO/STO system, and compare with the n-type model discussed so far in IL nickelate thin-films, with an all-cation element map including Sr, using STEM-EELS. Our studies are on a STO/PSNO/STO sample with p-type interface obtained through topotactic reduction with CaH$_{2}$ and an AlOx/PSNO/STO sample with n-type interface obtained by in-situ chemical reduction from an Al overlayer. This in-situ reduction using aluminium has been previously reported as an efficient technique to synthesize superconducting IL nickelate samples \cite{wei2023solid,wei2023superconducting}. See methods for more details on the synthesis. Considering the presence of Sr in the substrate, and it being the dopant, mapping the Sr gives advanced insights into the interface in comparison with previous studies, where an element map of Sr is absent \cite{goodge2023resolving, osada2020superconducting}. \textbf{Figure \ref{fig_STEM_EELS}} displays the interface comparison using STEM-EELS, between the novel chemically sharp p-type interface and the B-site intermixed n-type. Figure \ref{fig_STEM_EELS}a shows the STEM-high angle annular dark field (HAADF) image of the SC STO-capped PSNO/STO sample reduced using CaH$_{2}$ \cite{gutierrez2024towards}. Figure \ref{fig_STEM_EELS}b shows the intensity profile of the combined element map in Figure \ref{fig_STEM_EELS}c. The isolated maps of each element from their respective EELS edges is given in Figure \ref{fig_STEM_EELS}d. A zoom into the substrate interface region is given in Figure \ref{fig_STEM_EELS}e-g, which evidences the chemically sharp interface, with a Sr-layer separating the Ni and Ti on either sides, as indicated by the arrow. The Sr-layer interfacial layer was also observed in the perovskite phase before reduction, as shown in Supporting Information, Figure S1. This interface is chemically sharp and essentially avoids the B-site intermixing. Moreover, we observe Ni/Pr ratio slightly larger than 1 near both top and bottom interfaces as highlighted in Figure \ref{fig_STEM_EELS}b. This can be read as the distribution of Sr, and hence the distribution of holes. This will be discussed in detail in a later section on spatial mapping of hole-distribution. We also observe the same chemically sharp interface in another SC uncapped PSNO/STO sample prepared the same way (Supporting Information, Figure S2). In contrast, we observe the B-site intermixed interface in our in-situ reduced PSNO/STO sample, shown in Figure \ref{fig_STEM_EELS}h-j. The element map intensity profile in Figure \ref{fig_STEM_EELS}j shows the strong Ni/Ti intermixed interface, as indicated by the arrow, with the neighboring layers on either side being Pr. 
%Such an A-site intermixing might have been present in samples considered for previous studies, but couldn't be observed because of not having the map of Sr. In any case, 
Although this specific sample is not SC, this n-type interface is essentially similar to those reported up to now \cite{goodge2023resolving,krieger2022synthesis,yang2023thickness}. Similar results are found on another in-situ reduced sample with the same n-type interface, with resistivity plot showing a SC transition is given in Supporting Information, Figure S3. From now on we will refer to the chemically sharp interface as p-type and the B-site intermixed as n-type due to their respective interface polarity.         

\begin{figure*}[ht!]
  \centering
  %[width=7in]
  \includegraphics[width=\textwidth]{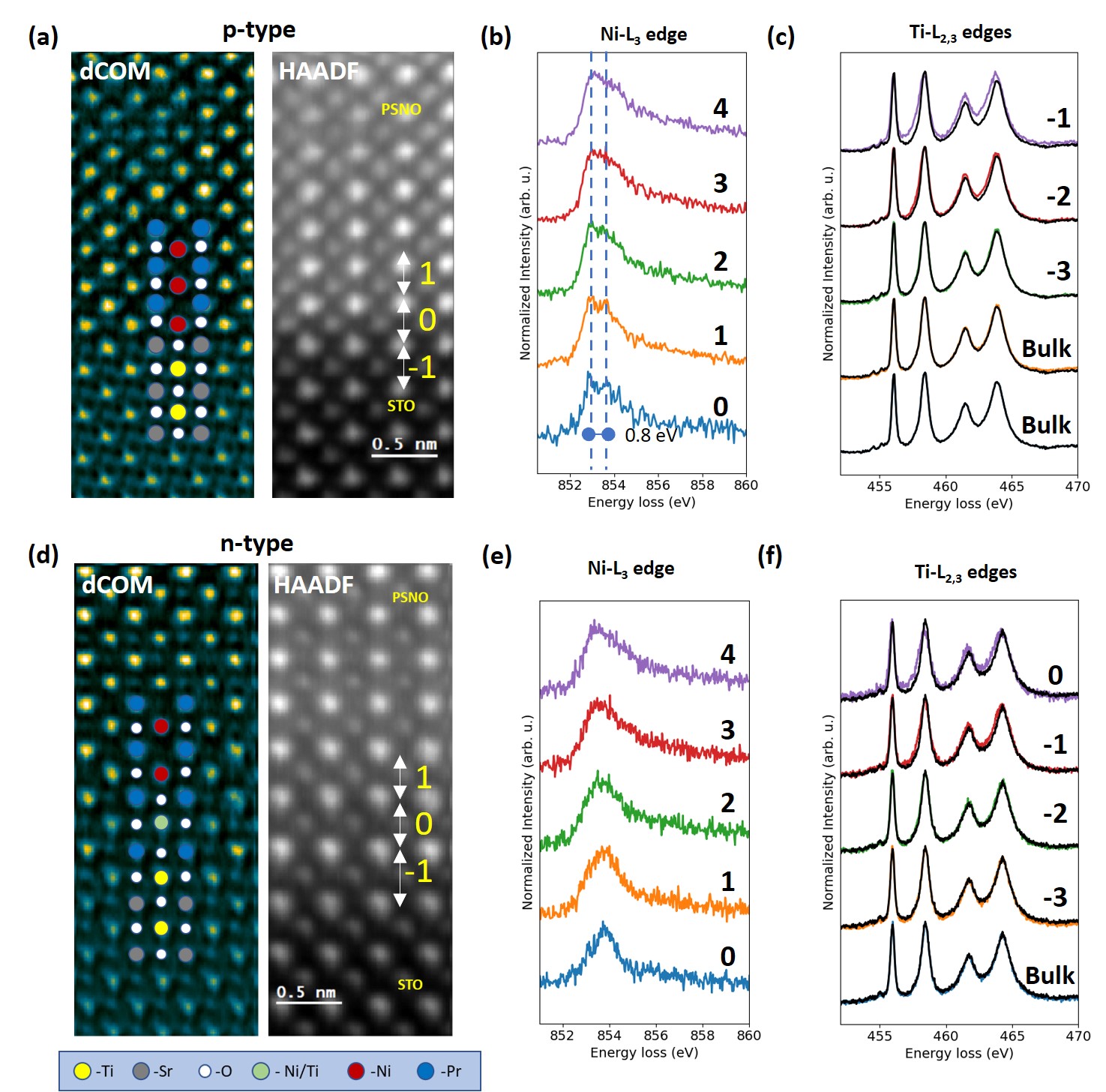}
  \caption{ 4D-STEM and Monochromated EELS analysis of SC STO-PSNO-STO and Al-PSNO-STO samples near the interface. (a) 4D-STEM dCOM and HAADF image with the interface model of STO-PSNO-STO thin film. (b) Ni-L$_{3}$ edge EELS at the u.c. as numbered from 0 to 4 near the interface. (c) Ti-L$_{2,3}$ edge EELS at the u.c. as numbered from -1 to -3 near the interface, and from bulk STO. One bulk is around 4 nm from the interface, and other 6nm, they are similar. (d) 4D-STEM dCOM and HAADF image with the interface model of Al-PSNO-STO thin film. (e) Ni-L$_{3}$ edge EELS at the u.c. as numbered from 0 to 4 near the interface. (f) Ti-L$_{2,3}$ edges EELS at the u.c. 0 to -3 near the interface, and from bulk STO. Notably, the STO-PSNO-STO sample do not have Ni$^{2+}$ at the interface, rather a doublet feature might be arising from orbital effects of the pyramidal Ni at the interface. There is a clear Ni$^{2+}$ at the interface u.c. in the Al-PSNO-STO sample. In both samples, there is no clear signature of Ti$^{3+}$.
}
   %\label{fig:boat2}
  \label{fig_4D_STEM_EELS}
\end{figure*}

To evaluate the associated structural and electronic evolution of these two interfaces, we employed 4D-STEM and monochromated EELS with atomic resolution. \textbf{Figure \ref{fig_4D_STEM_EELS}}a shows the divergence of the center of mass (dCOM) obtained by 4D-STEM measurements, which qualitatively approximates to a projected charge-density image and indicates the distribution of oxygen across the interface, along with the associated HAADF image of this p-type interface. Here, the final STO u.c. labelled as u.c. -1 in HAADF, includes Ti in an octahedral coordination, as in the bulk STO. The interface u.c. (defined as u.c. 0) hosts Ni in a pyramidal coordination, and with a highly asymmetric A-cation geometry, with Sr from the STO side and Pr from the PSNO side. From u.c. 1, Ni starts to be in square-planar coordination with symmetric Pr on either side. This essentially evidences a direct transition from the perovskite STO to the IL PSNO, without having any structural reconstruction at the interface. This makes the interface highly polar, with SrO[0 charge] - NiO$_2$[-3 charge] - Pr[+3 charge], which has the potential to host a strong electric field. The electronic structure of this interface is evaluated from the u.c. resolved EELS fine structure shown in Figure \ref{fig_4D_STEM_EELS}b,c. The Ni-L$_{3}$ edge EELS shown in Figure \ref{fig_4D_STEM_EELS}b reveals significant differences between Ni at the interface (u.c. 0) and those above. All the Ni exhibits primarily Ni$^{1+}$ character indicated the intense low energy edge, but with a doped shoulder at a higher energy loss, similar to the ones reported for a doped Nd$_{0.8}$Sr$_{0.2}$NiO$_{2}$ \cite{goodge2021doping}. However, one significant difference is observed for the Ni in u.c. 0, which is in a pyramidal coordination. Its Ni-L$_{3}$ edge exhibits a doublet feature, which could come from additional hole or some orbital ordering at the interface. Along with this, the u.c. resolved Ti-L$_{2,3}$ edges shows Ti$^{4+}$ character throughout, but with a more broadened spectral shape in the u.c. -1. Since it is a p-type interface, this small broadening is not expected to arise from mobile electrons contributed by the Ti 3d conduction band. This could arise from localized state and defect, or from the influence of the strong electric field at the interface, as it has been previously shown to broaden the Ti-L$_{2,3}$ edges at a nickelate-titanate interface \cite{grisolia2016hybridization}. Such an influence of the interface polarity will be discussed in detail in the coming section.

On the other hand, Figure \ref{fig_4D_STEM_EELS}d-f shows a similar characterization done on the n-type interface, which displays a very different structural and electronic landscape. Figure \ref{fig_4D_STEM_EELS}d shows the dCOM map and the HAADF image of the interface region in this system. As shown, the final STO u.c. is labeled as -1 in the HAADF image, with Ti in octahedral coordination, and with an asymmetric A-cation distribution with Sr from the STO side and the Pr on top. The u.c. 0 consists of the Ni/Ti mixed layer, these cations being located in octahedral coordination and symmetric Pr-sites on either side. The u.c. 1 hosts Ni in a pyramidal coordination, and a symmetric A-site with Pr on either sides. The NiO$_2$ square planar begins from the u.c. above it. Such an interface encompassing the perovskite-intermixed nickelate layer is less polar, and it is expected to have a smaller interface electric field compared to the p-type discussed before. The u.c.-resolved Ni-L$_{3}$ edge EELS shown in Figure \ref{fig_4D_STEM_EELS}e indicates a strong Ni$^{2+}$ character for the Ni at the u.c. 0, and a bit weaker Ni$^{2+}$ character for the Ni at u.c. 1. Ni exists in octahedral coordination in the former, and in pyramidal in the latter. In the following unit cells, the Ni exhibits strong Ni$^{1+}$ character, but not with a strong hole-doped shoulder as in Figure \ref{fig_4D_STEM_EELS}b. Also, as shown in Figure \ref{fig_4D_STEM_EELS}f, the Ti-L$_{2,3}$ edges does not exhibit notable differences from the bulk to the intermixed interface u.c. 0, and this indicates the absence of Ti$^{3+}$ as previously suggested for such an interface \cite{goodge2023resolving}. The absence of any spectral broadening of this Ti-L$_{2,3}$ edges at the interface is also consistent with a weaker electric field here. 

\subsection{Interface polarity}
  
\begin{figure*}[ht!]
  \centering
  %[width=7in]
  \includegraphics[width=\textwidth]{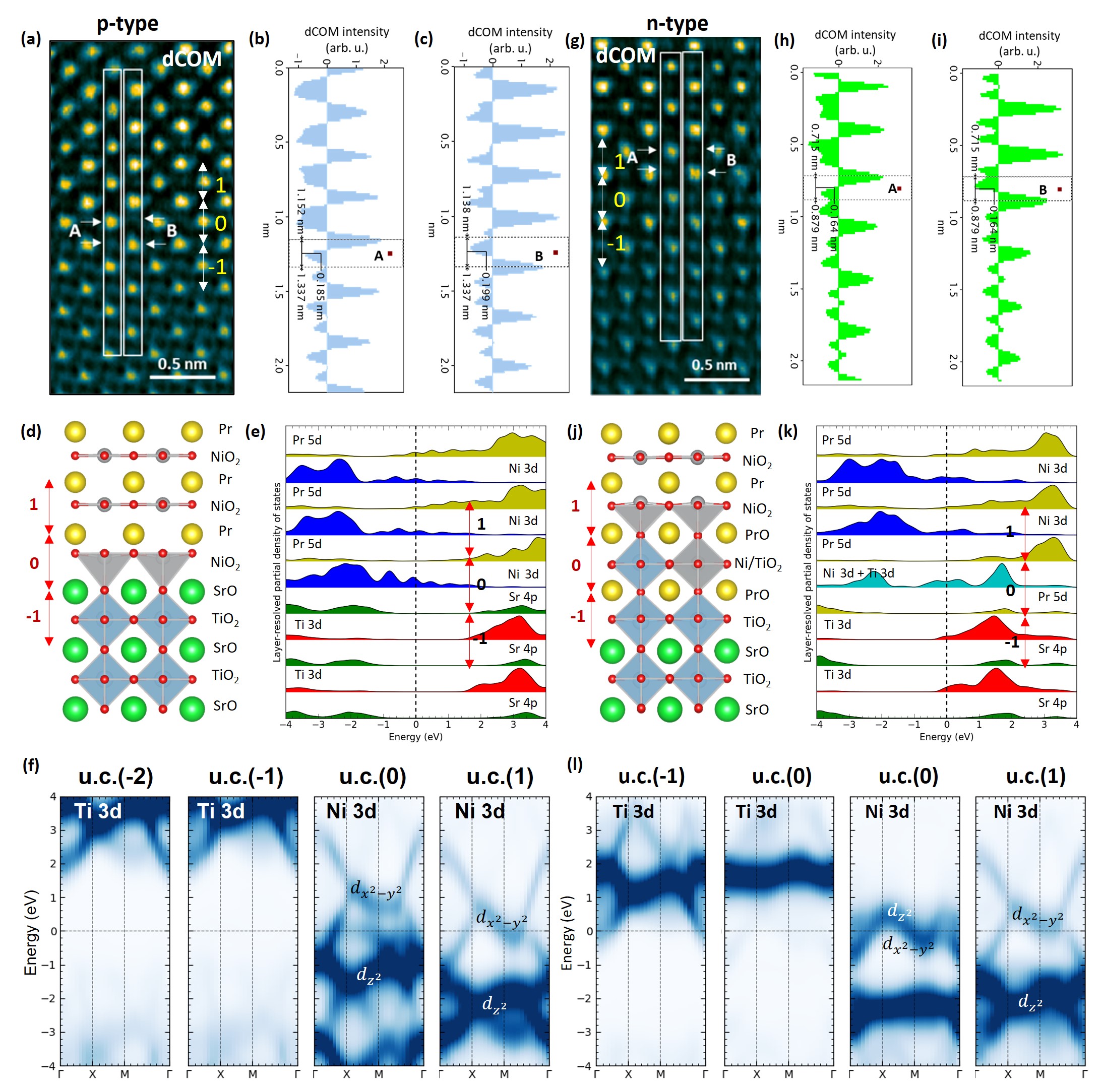}
  \caption{ Polar displacements at the interface and electronic structure of p-type (a-f) and n-type (g-l) interfaces. (a-c) 4D-STEM dCOM and the intensity profiles showing polar-like displacement of oxygen at the p-type interface. (d) DFT relaxed interface structure, reproducing the similar displacements of oxygen atoms as in the 4D-STEM dCOM. (e)  Layer resolved PDOS of the structural model in (d). (f) k-resolved PDOS of Ti 3d and Ni 3d orbitals from u.c. -2 to 1. (g-i) 4D-STEM dCOM and the intensity profiles indicating the absence of any relative displacement of oxygen at the n-type interface. (j) DFT relaxed interface structure, reproducing relatively smaller displacements between atoms at the interface as in the 4D-STEM dCOM. (k)  Layer resolved PDOS of the structural model in (j). (l) k-resolved PDOS of Ti 3d and Ni 3d orbitals from u.c. -1 to 1.}
   %\label{fig:boat1}
   \label{fig_polar_displacements}
\end{figure*}

The spectroscopic signatures in Figure \ref{fig_4D_STEM_EELS} thus indicate that, compared to the n-type interface, the p-type interface is highly polar and is expected to host a strong electric field. To make this more evident, we analyzed the atomic displacements at this interface with 4D-STEM dCOM data, also comparing with ab-initio calculation models of such an interface. \textbf{Figure \ref{fig_polar_displacements}}a-c shows such an analysis of the atomic displacements. The Ni-Sr distance is labeled as A, and the O-Sr distance is labeled as B for the u.c. 0. Corresponding to the arrows labeled in the dCOM map in Figure \ref{fig_polar_displacements}a, the distance A (Ni-Sr) is estimated as 0.185 nm in Figure \ref{fig_polar_displacements}b and the distance B (O-Sr) is estimated as 0.199 nm in Figure  \ref{fig_polar_displacements}c. This indicates a relative difference of 14 pm. Such atomic displacements are qualitatively observed in the relaxed structural model in Figure \ref{fig_polar_displacements}d, that further corroborates the strong interface polarity and the emergent electric field here. The layer-resolved partial density of states (PDOS) shown in Figure \ref{fig_polar_displacements}e, also clearly identifies it as the p-type interface with the Fermi level close to the valence band.  The k-resolved PDOS shown in Figure \ref{fig_polar_displacements}f indicates that the Ti 3d bands in u.c. -2 and u.c. -1  are very far from the Fermi-level, at energies larger than 2 eV. This is consistent with the absence of any strong signature of Ti$^{3+}$ near the interface. As expected from the EELS analysis, the pyramidal Ni at the u.c. 0 has a very different electronic structure in comparison with the square-planar Ni at the u.c. 1. The Ni 3d bands in the pyramidal Ni at u.c. 0 are under the influence of a strong electric field due to the high polarity mismatch as discussed before. On combining with the PDOS shown in Supporting Information, Figure S4, it seems that the occupied 3d$_{z^{2}}$ bands get shifted to higher energy, as it was also discussed in \cite{bernardini2020stability}. This modified orbital hierarchy where 3d$_{z^{2}}$ bands get closer to the Fermi level, could potentially be causing the doublet of Ni L$_{3}$ edge observed in EELS fine structure. On similar grounds, a complex orbital rearrangement has been reported in a related LaNiO$_2$/LaGaO$_3$ interface with a pyramidal Ni layer \cite{ortiz2021superlattice}. There the extra holes were getting accommodated with a multi-orbital character at this site, instead of getting evenly distributed across the layers \cite{ortiz2021superlattice}. The Ni 3d bands in the square-planar u.c. 1 experience a lower electric field, as can be seen in the k-resolved PDOS plot for u.c. 1. It is quite similar to the Ni 3d bands in the bulk IL as reported in previous studies \cite{geisler2021correlated}, and also for a single u.c. PrNiO$_2$, as shown in Supporting Information, Figure S5.           

In comparison, the n-type interface does not have any strong atomic displacements as shown in Figure \ref{fig_polar_displacements}g-i. This matches very well with the ab-initio calculation of the relaxed interface model shown in Figure \ref{fig_polar_displacements}j. The n-type nature of this interface is confirmed by the layer-resolved density of states shown in Figure  \ref{fig_polar_displacements}k, where the Fermi level is closer to the conduction band. This also goes in hand with the previous theoretical studies \cite{geisler2021correlated}. Notably, from the k-resolved PDOS shown in Figure \ref{fig_polar_displacements}l, the Ti 3d states in the STO seem to be located directly above, and cross the Fermi level a bit (u.c. -1), in contrast to the p-type interface. Interestingly, the Ti 3d bands in the u.c. 0, that is the Ni/Ti mixed u.c., show a bit of difference with Ti 3d in STO, and clearly do not have any bands near or crossing the Fermi level. This is consistent with the absence of any Ti$^{3+}$ inferred from EELS at the interface. On the other hand, in the octahedra at this u.c. 0, the Ni 3d$_{z^{2}}$ becomes unoccupied, in line with the demonstration of Ni$^{2+}$ at this u.c. by EELS. As already reported in the previous studies \cite{geisler2021correlated,geisler2020fundamental} and the PDOS shown in Supporting Information, Figure S4, the Ni 3d$_{x^{2} - y^{2}}$ stays near the Fermi-level, but with reduced bandwidth. Interestingly, the Ni 3d bands in the pyramidal u.c. 1 seem very different from the Ni 3d bands in the pyramidal u.c. 0 of the p-type interface in Figure \ref{fig_polar_displacements}f. However, it is similar to the Ni 3d bands in the square-planar u.c. 1 of the p-type interface. This difference might arise from the fact that the pyramidal Ni (u.c. 0) of the p-type is under the influence of the strong electric field since the interface is highly polar. However, such a strong electric field is absent for the n-type interface, and thus the pyramidal Ni in u.c. 1 renders an orbital energetic hierarchy similar to the square-planar Ni \cite{raji2024valence}. From these results, one can understand that the two interfaces have a very different electronic nature that might influence the electronic structure of neighbouring layers. However, it is to be noted here that in both cases, the influence of the interface vanishes beyond 2-3 unit cells into the nickelate layer.      

\subsection{Spatial hole distribution}
\begin{figure*}[ht!]
  \centering
  \includegraphics[width=\linewidth]{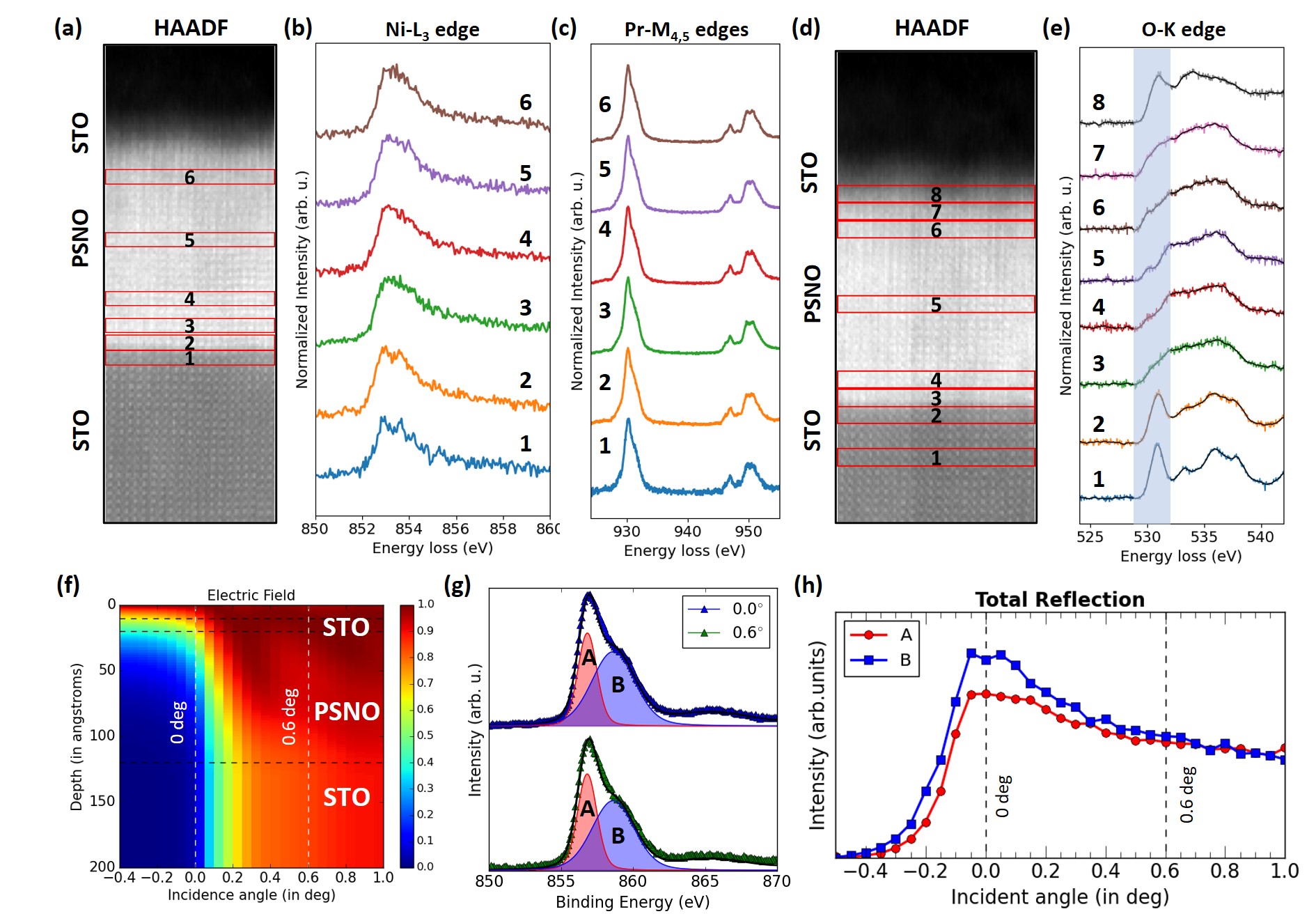}
  \caption{ Macroscopic mapping of the electronic structure of SC STO-PSNO-STO using STEM-EELS and Total Reflection(TR)-HAXPES.  (a) HAADF image of STO-PSNO-STO thin film.  (b,c) EELS Ni-L$_{3}$ and Pr-M$_{4,5}$ edges at different regions of the thin film shown in (a). (d) HAADF image of STO-PSNO-STO thin film. (e) EELS O-K edge at different regions shown in (d). (f) The Electric field at the total reflection condition using YXRO code \cite{yang2013making}, for an STO/PSNO/STO system with around 1 nm of carbon on top. (g) Ni 2p core level PES at two different angles around the total reflection condition. (h) Rocking curve with the angle of the two components of Ni-2p at the total reflection condition. The Ni shows more hole doping near the top and bottom interface.
}
  %\label{fig:boat1}
  \label{fig_macro_map}
\end{figure*}

 One hallmark of the IL nickelate superconductivity is its SC dome upon varying the hole doping from 10 to 25 $\%$ \cite{li2020superconducting}. It is thus important to study the distribution of this doped holes throughout the nickelate layer, and in combination with local structural and chemical effects, this could be one key factor dictating superconductivity in this system. Here using monochromated EELS over the whole thin film, and with depth-resolved total reflection(TR)-HAXPES, we carried out a spatial mapping of the charge/hole distribution in a superconducting STO-capped PSNO/STO sample with the p-type interface. \textbf{Figure \ref{fig_macro_map} }a-e shows the evolution of Ni-L$_{3}$, Pr-M$_{4,5}$, and O-K EELS edges across the whole thin film of the superconducting STO-PSNO-STO sample. The Ni-L$_{3}$ edge shown in Figure \ref{fig_macro_map}b forms a representative map  of the hole distribution, as the higher energy loss shoulder emerges with Sr-doping as reported in previous studies with EELS and XAS \cite{goodge2021doping,rossi2021orbital}. Notably, this feature is spatially varying and it emerges near the top and bottom interfaces. Especially in region 4, the hole-doped signature is weaker. To further corroborate this, as shown in Supporting Information Figure S6, we did a separate spatial map of the ratio of a pure Ni$^{1+}$ component (from reference \cite{raji2023charge}) and Ni -hole doped component in this thin-film, and it shows the same trend. The Ni-hole doped component exhibits an increase concentration near the top and bottom interface, except the middle. But it always stays within the superconducting dome \cite{wang2023experimental}, as represented in Supporting Information Figure S6b. This goes along with the Ni/Pr off ratio seen in the element map in Figure \ref{fig_STEM_EELS}b, which suggested that the doped 20$\%$ Sr gets distributed near both interfaces. Thus this spatial difference in the spectral shape of Ni-L$_{3}$ edge can be directly coupled with the Sr-distribution in the thin film, and having the ideal hole doping coupled with the perfect IL structure at the regions near the interfaces could be core requirement for IL nickelate superconductivity. In addition, as shown in Figure \ref{fig_macro_map}c, the Pr-M$_{4,5}$ edges does not show any difference spatially and this is expected because it is quite insensitive to local electronic or chemical changes \cite{raji2023charge}. On the other hand, the spatial mapping of the O-K edge shown in Figure \ref{fig_macro_map}e, displays strong differences throughout the thin film. The spectra from the regions 3 and 6,7 that are near the bottom interface and top of the thin film exhibit a reduced pre-edge and it is representative of the spectra of a doped Ni$^{1+}$ \cite{goodge2021doping}. For the region 5, that is in the middle of the thin film, the O-K edge exhibits a stronger pre-edge and a spectral feature representative of a defect/non-stoichiometric region as reported in previous studies \cite{raji2023charge, goodge2021doping}. 

To address the uniformity of such an epitaxial distribution of holes over the whole sample, we employed a macroscopic depth resolved analysis using TR-HAXPES. In this method, in addition to obtaining a bulk sensitive photoemission spectra, one can tune the incidence angle in a way to have selective probing of the top or bulk of the thin film with good spatial resolution \cite{cambou2021depth}. Figure \ref{fig_macro_map}f shows the depth-resolved simulated electric field that will be generated in this STO/PSNO/STO system from x-rays when one varies the incidence angle in photoemission near the total reflection angle. Depending on the depth of the electric field, the photoemission signal will be enhanced up to that depth. This is the principle of TR-HAXPES, and it is given in detail in Methods. Figure \ref{fig_macro_map}g shows the Ni 2p$_{3/2}$ photoemission spectrum for two different incident angles. It shows a Ni$^{1+}$ character, but with a component at a higher binding energy labeled as B, representative of the hole doping. On par with the EELS analysis, the B-component gets stronger at 0 deg, that is when the electric field enhances the photoemission signal from the top of the nickelate layer. At a higher angle 0.6 deg, when the signal probes the entire nickelate layer, the relative intensity of the B-component becomes smaller compared to that at 0 degrees. When compared with the rocking curve in Figure \ref{fig_macro_map}h shows that the B-component is relatively stronger for the incidence angles between -0.2 and 0.2 deg, that is when the electric field is stronger at the top of the nickelate layer as illustrated in Figure \ref{fig_macro_map}f. It has to be considered that some of the regions between the middle of the nickelate layer to the bottom interface lacks Sr-doping as evidenced by EELS. As shown in Figure \ref{fig_macro_map}f, on angles higher than 0.25 deg, the electric field is almost homogeneous and strong in the whole nickelate layer. This indicates that up to 0.25 deg, we can selectively enhance photoemission from top of the nickelate layer, after that, the signal gets averaged out in the whole layer. This gives us the constrain that we can only isolate the signal from top nickelate region, and not the middle or bottom region from the whole signal. In this way, the photoemission signal over the whole sample at higher incidence angles, the signal from the Sr-rich and Sr-poor regions gets averaged out. Thus, the hole-doping component B, emerging from Sr-doping gets almost the same intensity as the main Ni$^{1+}$ component A. This concomitantly proposes more doped holes in the thin film top than in the middle. On combining with the microscopic results, this can be read as doped Sr-getting bit enriched near both interfaces, and such a trend might be present throughout the sample as probed macroscopically by TR-HAXPES. 
\section{Conclusion}
It is now demonstrated that both p-type and n-type interfaces can form in IL nickelate thin films. While from previous works, the n-type is superconducting, here we show the p-type is superconducting as well. It directly questions the role of the interface as a primary factor in controlling superconductivity. In particular, our u.c. resolved EELS fine structure and ab-initio calculations indicates that after 2-3 unit cells, the interface influence is weaker on the nickelate. On the other hand, intrinsic doping arising from Sr, along with self-doping effects from local off-stoichiometry are prone to modulate the hole-doping within the layer. Along with this, different interface reconstructions can act as a secondary hole source, however, the occurrence of superconductivity is not bound to a specific interface model. These results unravels the non-universality of the interface in SC IL nickelates.

\section{Experimental Section}

\subsection{Sample preparation}
Perovskite Pr$_{0.8}$Sr$_{0.2}$NiO$_{3}$ nickelate thin films of thicknesses around 8 nm were grown in a PLD system equipped with a KrF excimer laser (248 nm) focused onto the target \cite{gutierrez2024towards}. The laser pulse rate was fixed at 4 Hz. During the growth, oxygen was supplied in the PLD chamber yielding a background pressure up to 0.5 mbar. The laser fluence was 1.6 $\mbox{J}\,\mbox{cm}^{-2}$, with a laser spot size of $2\pm 0.2 \;\mbox{mm}^2$.  The substrate temperature was set at 640 $^{\circ}$C, and the oxygen pressure at 0.33 mbar. The films were cooled down to room temperature at a rate of 5 $^{\circ}\mbox{C}\,\mbox{min}^{-1}$ at the growth pressure. Single crystals of (001)SrTiO$_3$ (STO) were used as substrates. Prior to the growth, they were etched in buffered HF and annealed at 1000 $^{\circ}$C for 3 h to obtain stepped surfaces with TiO$_{2}$ termination. No further annealing under vacuum was carried out before the film growth. We sintered the PLD target from a mixture of Pr$_2$O$_3$ (99.99\%), Nickel(II) oxide (99.99\%) and  SrCO$_3$ with controlled cation stoichiometry ([Ni]/([Pr]+[Sr])=1.1; [Pr]/[Sr]=4) by a solid state reaction.  These mixtures were ground in an agate mortar and, after initial decarbonation at 1200 $^{\circ}$C for 12 h, pressed into pellets, and heated in a box furnace at 1300 $^{\circ}$C for 24 h. To get high-density PLD targets, the powders were reground and repressed, and then fired at 1300 $^{\circ}$C for further 24 h.  

Reduction of the perovskite phase thin films into the IL phase was carried out by two different methods: CaH$_{2}$ reduction and in-situ Al reduction. Regarding the CaH$_{2}$ reduction, the process was performed in evacuated glass tubes. The tubes were filled with 0.1 g of CaH$_2$ powder in an N$_2$-filled glovebox. Samples were wrapped in aluminum foil and inserted into the glass tubes.  The samples are separated from the CaH$_2$ powder (Alfa Aesar A16242) by means of a lump of glass wool. The tubes were evacuated by a rotary pump ($< 10^{-3}$ mbar) and sealed.  A horizontal tube furnace, ventilated with N$_2$ for increased temperature homogeneity was used for the process, with temperature accuracy $\pm 1\; ^{\circ}$C.  The reference temperature is measured at the center of the tube furnace, where the sealed ampoule is placed.  Heating and cooling rates in the oven were 5 $^{\circ}\mbox{C}\;\mbox{min}^{-1}$.

In the case of in-situ reduced Al/PSNO/STO samples, the deposition of 3.5 nm of aluminum metal overlayer was carried out inside sputtering chamber (from PLASSYS) at temperature of 380 $^{\circ}\mbox{C}$ and pressure of 6 × 10$^{-4}$mbar. The argon flow was set to 5.2 sccm,the current was set to 15 mA, power at 5 W and voltage at 320V. Before Al depositon the target was pre-sputtered for 10 minutes to remove possible oxidized layer. After Al deposition, the samples are kept at same temperature as deposition (380$^{\circ}$C) for 2h before cooling. The heating and cooling rates were $\pm$ 10 $^{\circ}\mbox{C}$/min.
\subsection{High resolution STEM-EELS and 4D-STEM}
Cross-sectional transition electron microscopy (TEM) lamellae were prepared using a focused ion beam (FIB) technique at C2N, University of Paris-Saclay, France). Before FIB lamellae preparation, around 20 nm of amorphous carbon was deposited on top for protection. The HAADF imaging and 4D-STEM were carried out in a NION UltraSTEM 200 C3/C5-corrected scanning transmission electron microscope (STEM). The experiments were done at 200 keV with a probe current of approximately 14 pA and convergence semi-angles of 30 mrad. A MerlinEM (Quantum Detectors Ltd) in a 4×1 configuration (1024 × 256) has been installed on a Gatan ENFINA spectrometer mounted on the microscope \cite{tence2020electron}. The EELS spectra are obtained using the full 4 × 1 configuration and the 4D-STEM by selecting only one of the chips (256 × 256 pixels). For 4D-STEM, the EELS spectrometer is set into non-energy dispersive trajectories and we have used a 6-bit detector mode that gives a diffraction pattern with a good signal-to-noise ratio without compromising much on the scanning speed. 
The monochromated EELS have been done using a NION CHROMATEM STEM at 100 keV with a probe current of approximately 30 pA, convergence semi-angles of 25 mrad and an energy resolution of around 70 meV. The EELS detection was also done with a MerlinEM in a 4×1 configuration (1024 × 256) that has been installed on a Nion IRIS spectrometer mounted on the microscope.

\subsection{Ab-initio Calculations}
The first principles calculation were performed using the density functional theory \cite{kohn1965self} as implemented in the Quantum ESPRESSO package \cite{giannozzi2009quantum,giannozzi2017advanced}. The exchange-correlation functional was approximated by the generalized gradient approach \cite{perdew1996generalized}. Planewave cutoffs of 51 and 526 Ry were used basis-set and charge density expansions, respectively. Structural relaxations and electronic structure calculations were done with a Hubbard U parameter of 4eV for Ni and Ti elements.The p-type and n-type interface models were created in superlattice geometry with in-plane $\sqrt{2}$a × $\sqrt{2}$a and out-of-plane c. This gives two transition metal atoms per layer. By constraining this to the substrate STO, it gives in-plane lattice parameters of a = b = 5.52 \AA\ . To understand the band folding that might arise, we also did the calculation for a single u.c. of PrNiO$_{2}$ with in plane 1a x 1a and and in-plane $\sqrt{2}$a × $\sqrt{2}$a. This is shown in Supporting Information, Figure S5. 

\subsection{HAXPES measurements}
The measurements were carried out at the GALAXIES beamline at the SOLEIL synchrotron \cite{rueff2015galaxies} on the HAXPES endstation \cite{ceolin2013hard} using a photon energy of 6000 eV, with varying incidence angles from -1$^\circ$ to 80$^\circ$. The electric field at TR-condition was simulated from the YXRO code \cite{yang2013making}. The synchrotron operated with a ring current of 450 mA, giving an intensity of 3.4 × 10$^{13}$ photons/s at 3000 eV, which was then reduced using a built-in filter to 5\% of the original intensity. The photoelectrons were detected using a SCIENTA Omicron EW4000 HAXPES hemispherical analyzer, and a Shirley background \cite{shirley1972high} was removed before fitting the core levels spectra.

\medskip
\section{Supporting Information} \par %Please delete the Suppporting Information statement if it is not applicable. Please supply Supporting Information in another file. Supporting information should not be provided in .tex format
Supporting Information is available from the Wiley Online Library or from the author.
% Acknowledgements

\medskip
\section{Acknowledgements} \par %delete if not applicable))
A.R. acknowledges financing from LABEX NanoSaclay and H2020 for the doctoral funding. A.G.L. acknowledges financial support through a research grant from the Next Generation EU plan 2021, European Union. Nion CHROMATEM at LPS Orsay and the FIB at C2N, University of Paris-Saclay was accessed in the TEMPOS project framework (ANR 10-EQPX-0050). This work is supported from the framework of the joint ANR-RGC ImagingQM project (RGC,A-CityU102/23;ANR,ANR-23-CE42-0027). We acknowledge SOLEIL Synchrotron for provision of beamtime under proposals 20221574 and 20231845. We thank Alberto Zobelli, Daniele Preziosi, and Danfeng Li for fruitful discussions. 

% References
\medskip

% Use the following code if you wish to generate your bibliography with BibTeX;
% replace the string "MSP-template" below with the name(s) of
% the BibTeX data base(s) you want to use.
% The resulting bibliography-output (the content of the .bbl file)
% must be pasted back into this file before submission.
% Please also include your BibTeX data base file(s) in your submission
% so that we can re-run BibTeX if necessary.
%
\bibliographystyle{MSP}
\bibliography{references}

% Figures/tables and captions
% Permission statements are required for all figures reproduced or adapted from previously published articles/sources. Please also ensure that all necessary permissions to reproduce images have been received
% Please remove these statements for original figures
\appendix
\section{Supporting Information}


\section*{Supplementary material}

Supporting Information for

# Unraveling p-type and n-type interfaces in Superconducting Infinite-Layer Nickelate Thin-films

Aravind Raji[1,2,⋆], Araceli Gutiérrez-Llorente[3], Dongxin Zhang[4], Xiaoyan Li[1], Manuel Bibes[4], Lucia Iglesias[4], Jean-Pascal Rueff[2,5,⋆], and Alexandre Gloter[1,⋆]

[1]Laboratoire de Physique des Solides, CNRS, Université Paris-Saclay, Orsay, 91400, France
[2]Synchrotron SOLEIL, L'Orme des Merisiers, BP 48 St Aubin, Gif sur Yvette, 91192, France
[3]Universidad Rey Juan Carlos, Escuela Superior de Ciencias Experimentales y Tecnología, Madrid 28933, Spain
[4]Laboratoire Albert Fert, CNRS, Thales, Université Paris-Saclay, 91767 Palaiseau, France
[5]LCPMR, Sorbonne Université, CNRS, Paris, 75005, France
⋆Corresponding authors. Emails: aravind.raji@universite-paris-saclay.fr, jean-pascal.rueff@synchrotron-soleil.fr, alexandre.gloter@universite-paris-saclay.fr

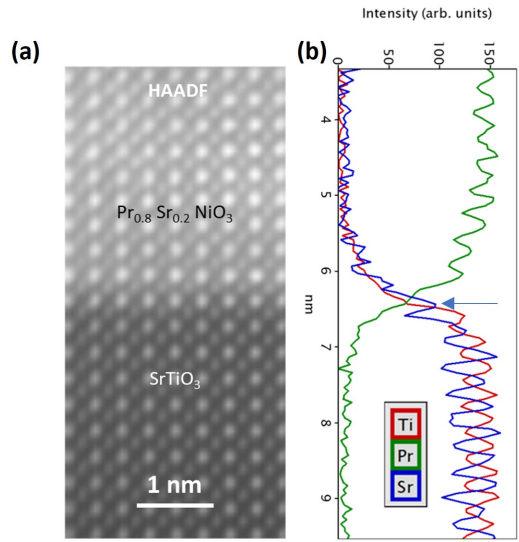

**Figure S1:** STEM-EELS interface elemental analysis of the doped parent perovskite $Pr_{0.8}Sr_{0.2}NiO_3$(PSNO3)/STO sample (a) HAADF image of the PSNO3-STO thin-film interface. (c) EELS element profile showing the Sr-interfacial layer, as observed for the reduced SC PSNO sample.

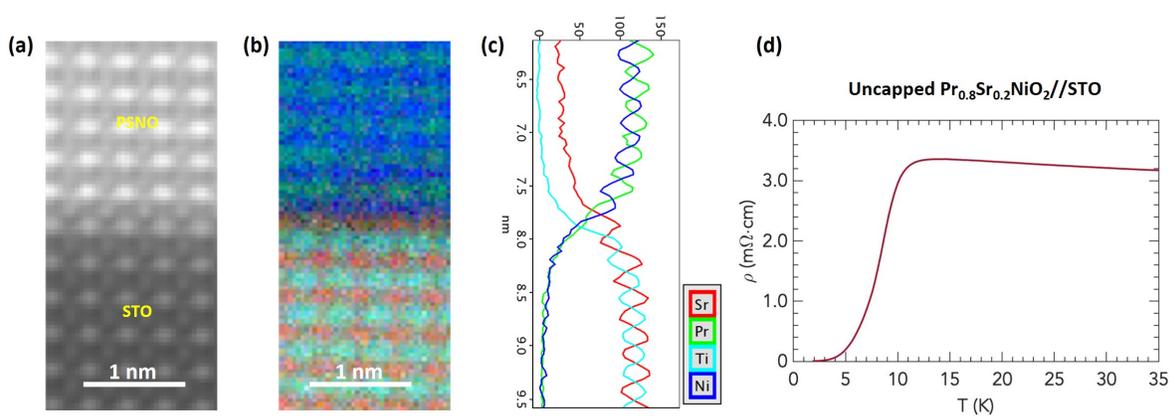

**Figure S2:** STEM-EELS interface elemental analysis of superconducting uncapped PSNO/STO sample with a p-type interface. (a) HAADF image of uncapped PSNO-STO thin-film interface. (b) EELS combined element map. (c) EELS element profile showing the chemically sharp p-type interface model. (d) The Resistivity vs Temperature plot showing the superconducting transition.

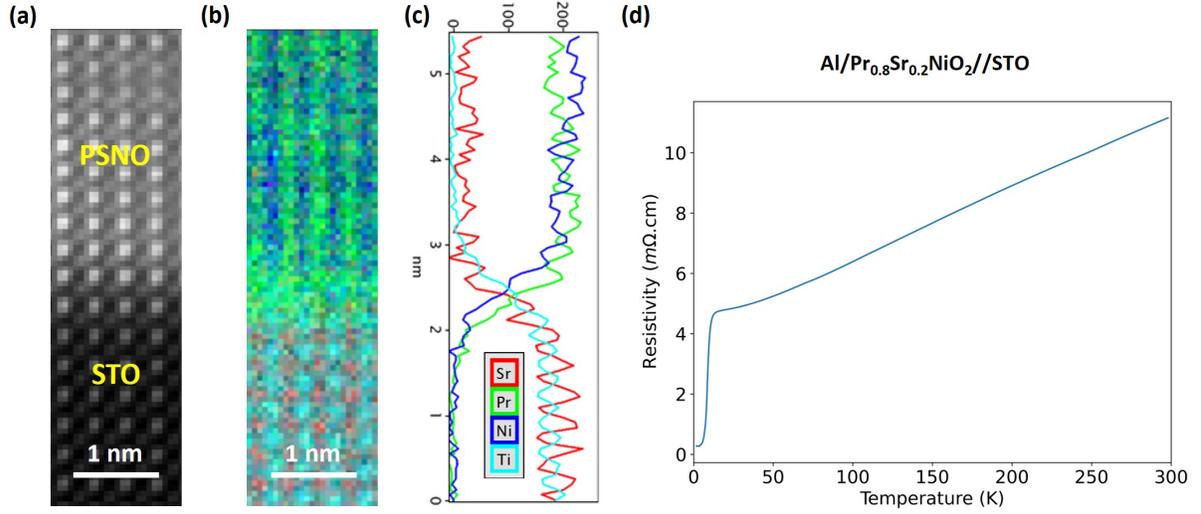

**Figure S3:** STEM-EELS interface elemental analysis of superconducting in-situ reduced Al-PSNO/STO sample with an n-type interface. (a) HAADF image of uncapped PSNO-STO thin-film interface. (b) EELS combined element map. (c) EELS element profile showing the chemically sharp p-type interface model. (d) The Resistivity vs Temperature plot showing the superconducting transition.

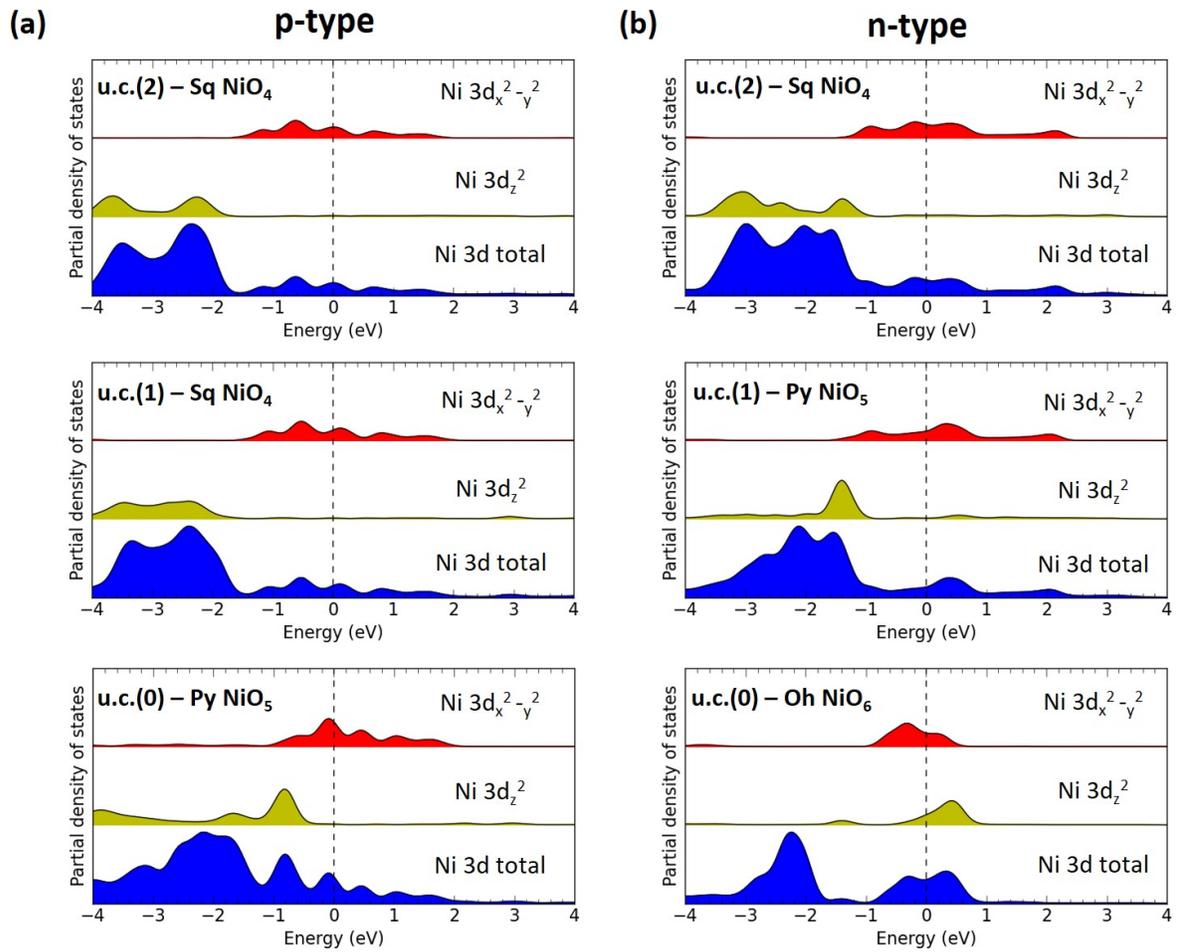

**Figure S4:** Partial density of states of Ni 3d orbitals from u.c. 0 to 2 in (a) p-type and (b) n-type interfaces

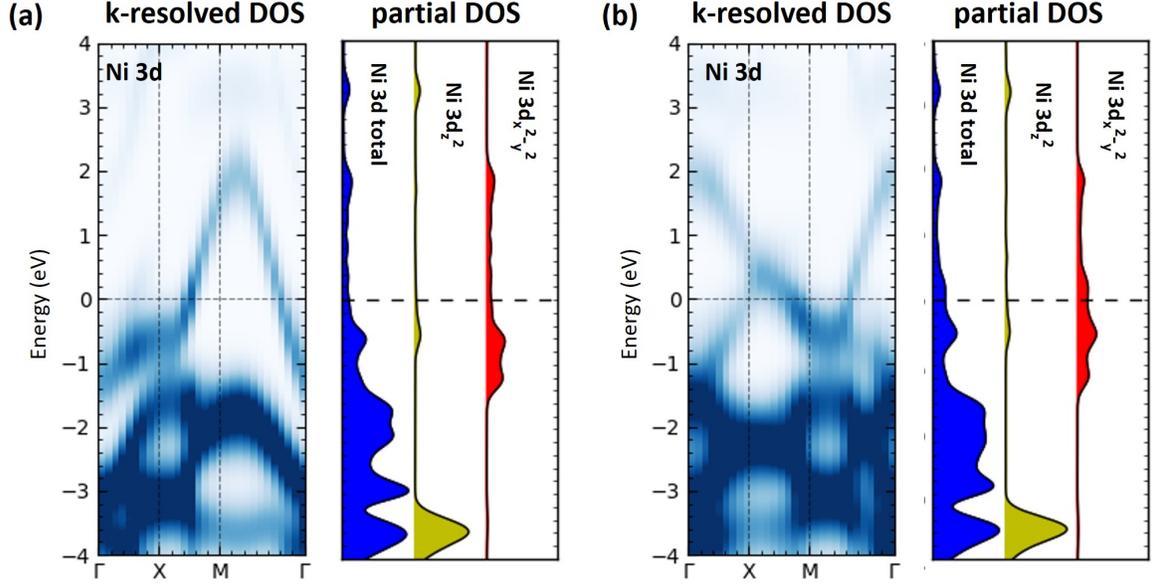

**Figure S5:** k-resolved and Partial density of states of Ni 3d orbitals of a single PrNiO2 unit cell with (a) in-plane 1 a x 1 a and out-of-plane 1 c (b) in-plane $\sqrt{2}a \times \sqrt{2}a$ and out-of-plane 1c lattice parameters.

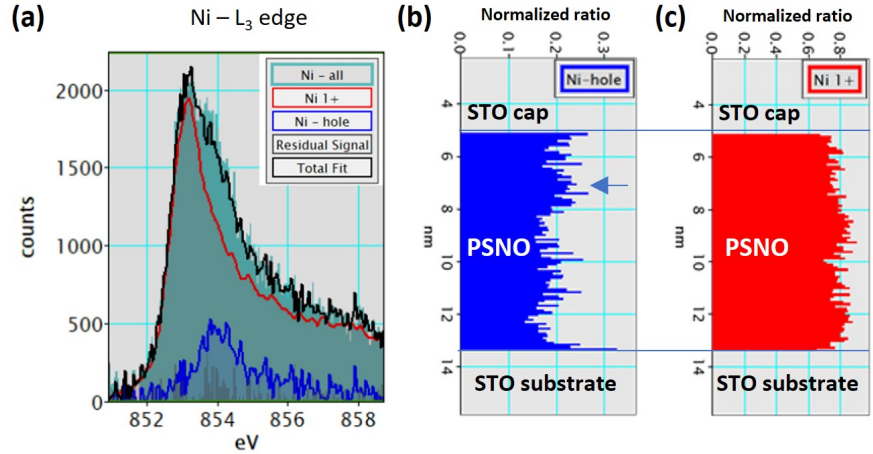

**Figure S6:** Spatial mapping of hole doping in the SC STO/PSNO/STO sample with the Ni-$L_3$ edge EELS. (a) Fitting of the EELS Ni-$L_3$ edge with the Ni 1+ component and the component from hole doping. (b) Spatial map of the hole doping component, that shows a higher ratio on the top of the thin-film, in the same region of Ni/Pr off stoichiometry as discussed in the main text. (c) Spatial mapping of Ni 1+ component that is more or less uniform, but with a bit higher ratio in the middle of the thin-film.


\end{document}